\documentclass[epj]{webofc}
\usepackage[varg]{txfonts}   
\woctitle{MESON2018 - the 15$^\textrm{th}$ International Workshop on Meson Physics}
\begin{document}
\selectlanguage{english}
\title{On the width of the $\Delta(1232)$ in $N\Delta$ and
$\Delta\Delta$ states}

\author{Jouni A. Niskanen\inst{1}\fnsep\thanks{\email{jouni.niskanen@helsinki.fi}} }

\institute{Helsinki Institute of Physics, University of Helsinki, Finland }

\abstract{
Due to the finite kinetic energy in the intermediate $N\Delta$
state the (internal) energy available for mesonic decay 
is decreased
and consequently the effective $N\Delta$ width is suppressed
in $NN$ scattering. The same can happen also in 
$\Delta\Delta$ case. Also the $N\Delta$ angular momentum 
suppresses the width as well, while the effect of the initial
$NN$ angular momentum is more subtle. The state dependence 
affects e.g. pion production observables and can also be
seen as the origin of $T=1$ ``dibaryons''.
}
\maketitle
\section{Introduction}
\label{intro}
\begin{figure}[hb]
\vspace{-1.5cm}
\centering
\sidecaption
\includegraphics[width=8cm,clip]{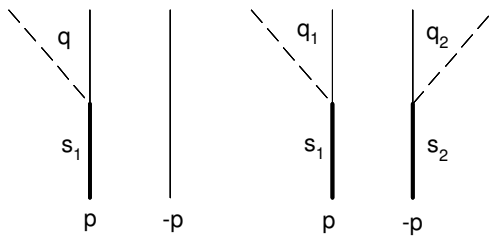}
\caption{Kinematics of the $N\Delta$ (left) and $\Delta\Delta$
(right) decays. The relative momenta $q_i$ between the pion 
and nucleon are given by $s_i = \sqrt{q_{(i)}^2+\mu^2} +
\sqrt{q_{(i)}^2+M^2}$ with $\sqrt{s_i}$ the internal energy 
of the $\Delta$.}
\label{Deltas3}       
\vspace{-1.5cm}
\end{figure}

$N\Delta$ configurations arise by coupled channels 
in various contexts (e.g.
pion production and absorption) as intermediate excited states
of the externally given $NN$ states. As such, below the nominal 
$N\Delta$ threshold the channel is naturally closed (virtual) 
and of finite range. Also at and above threshold the 
$N\Delta$ wave function is confined due to the finite pionic
decay width of the $\Delta$. In both cases the expectation value
of the kinetic energy would be finite in the channel. 
As seen from Fig.~\ref{Deltas3}, 
apparently the relative kinetic energy $E(p)$
does not participate in the decay as the invariants $s_i$ do,
but should rather be subtracted from the total overall 
energy~\cite{improv}. 

For nonzero angular momenta another aspect of kinetic energy, 
the centrifugal barrier, can obviously act as repulsion in the
$N\Delta$ channels strongly suppressing the corresponding wave
functions. This, in turn, causes strong state dependence to 
their effects~\cite{polarconf} further conveyed to observables.
It turns out that the state dependence goes also into the
widths giving each $N\Delta$ channel an \emph{effective}
width $\Gamma_{\rm eff}$. This leads to an improved agreement
with experiment in $pp \leftrightarrow d\pi^+$~\cite{improv}.

\section{$N\Delta$ states}
\label{single}
\begin{figure}[tb]
\vspace{-1cm}
\centering
\includegraphics[width=\columnwidth]{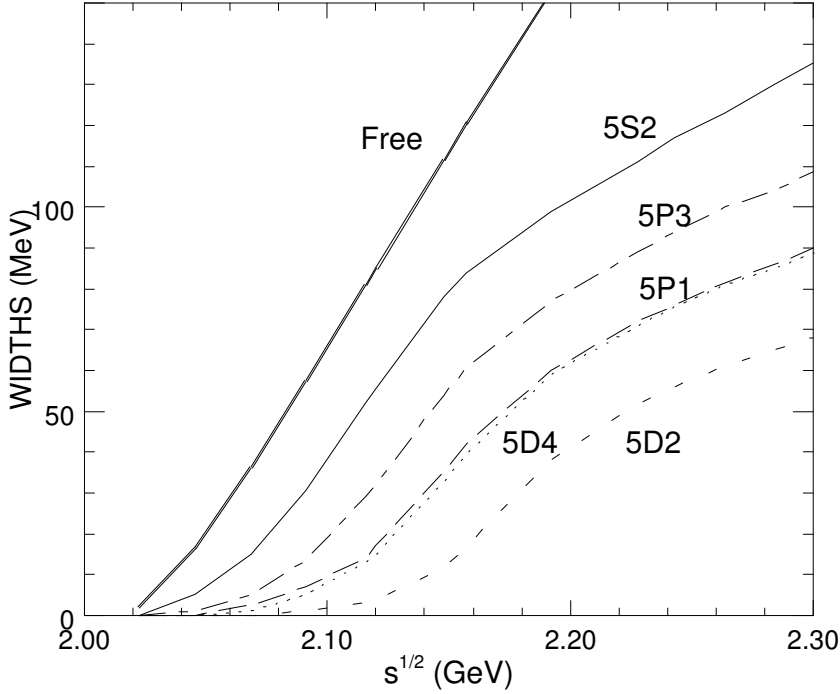}
\caption{The widths of a representative selection of $N\Delta$
states in $NN$ scattering: Curves as described in the text. The
free width is the thick line above the others. 
\label{widths}
}
\end{figure}

The effective decay width of the $N\Delta$ configuration, 
associated with $NN$ scattering, into the three-body final
state of Fig.~\ref{Deltas3} can be calculated explicitly 
as an average over kinematically allowed 
momenta~\cite{improv,prcnd}
\begin{equation}
\Gamma_{\rm 3,eff} = \frac{2}{\pi}\,
\frac{\int_0^{p_{\rm max}} |\Psi_{N\Delta}(p)|^2\,
\Gamma(q)\, p^2\, dp}
{\int_0^\infty |\Psi_{N\Delta}(r)|^2\, r^2\, dr} .
\label{gamma3}
\end{equation}
Here $\Psi_{N\Delta}(p)$ is the Fourier transform of the
appropriate partial wave component and $\Gamma(q)$ the free 
$\Delta \rightarrow N\pi$
width with $q$ as the relative $N\pi$ momentum. The kinematics
is determined by first subtracting the kinetic part from the
c.m.s. energy. The effect of this decrease of available decay
energy (and angular momentum barrier) is shown in 
Fig.~\ref{widths}. The half-width is used as an imaginary constant
potential input in the relevant $N\Delta$-channel Schr\"odinger
equation.
Clearly, there is a remarkable reduction from free width values 
for any $N\Delta$ angular momentum. The $^5S_2(N\Delta)$
configuration originates from $^1D_2(NN)$ scattering. The
$^5 D_2(N\Delta)$ (short dashes)
from the same initial state is much more
suppressed, as anticipated earlier. Further, an interesting 
interplay with $NN$ angular momenta shows up. The $^5 D_4(N\Delta)$
(from $^1 G_4(NN)$, dotted) is less suppressed. 
Similarly, $^5P_3(N\Delta)$ from $^3F_3(NN)$ (dash-dot)
is less suppressed than  $^5P_1(N\Delta)$ from $^3P_1(NN)$
(long dashes). In contrast the width of $^5D_0(N\Delta)$ from
$^1S_0(NN)$ is totally negligible below $E_{\rm lab}\approx
600$ MeV and above 800 MeV only about 15--20 MeV.
Apparently higher $NN$ centrifugal barriers
``push'' the baryonic state to $N\Delta$. This can be understood 
considering that the most influential range for the interaction 
is $\approx\! 1$ fm, and actually the $N\Delta$ wave function 
peaks around this range. 
Once the particles are at that distance, the
barrier is $\approx\! 40\times L(L+1)$ MeV and the loss of the
mass barrier $M_\Delta - M_N$
in transition is partly regained from the
diminished centrifugal barrier. However, the angular 
momentum barrier in the $N\Delta$ state is the dominant effect
and the $NN$ secondary. Still, it is remarkable that by these
arguments one gets the quantum numbers of
the isovector dibaryons of Ref.~\cite{yoko}
correctly and the mass values reasonably well from the above
rotational series~\cite{dibarmass}.

\section{$\Delta\Delta$ states}
In the case of two decaying particles it may be necessary to
specify how the lifetime is defined.
The decay rate for particles 1 and 2 with widths $\Gamma_1$
and $\Gamma_2$ starting from time zero is taken to be
$\Gamma_1 \exp(-\Gamma_1 t_1)\times \Gamma_2 \exp(-\Gamma_2 t_2)$.
The total transition probability at time $t$ is then (integrating
over different time orderings)
\begin{eqnarray}
P(t) = & \, \Gamma_1\Gamma_2 (\, \int_0^t e^{-\Gamma_1 t_1} dt_1
   \, \int_0^{t_1} e^{-\Gamma_2 t_2} dt_2 +
  \int_0^t e^{-\Gamma_2 t_2} dt_2 \, \nonumber
  \int_0^{t_2} e^{-\Gamma_1 t_1} dt_1 \, ) \\
  = & 1 - e^{-\Gamma_1 t} - e^{-\Gamma_2 t}   
      + e^{-(\Gamma_1 + \Gamma_2)t}  \, . 
\end{eqnarray}
and the survival probability
 $ 1 - P(t) = \exp(-\Gamma t)[\exp(-\delta t) + \exp(+\delta t)
 - \exp(-\Gamma t)]$ 
(with the notation $\Gamma = (\Gamma_1 + \Gamma_2)/2$
and $\delta = (\Gamma_1 - \Gamma_2)/2$).
So, the dominant part is consistent with
the decay width being the average $\Gamma$, or the single
width in the case $\Gamma_1 = \Gamma_2 = \Gamma$. In view of the 
kinematic results of Sec. \ref{single} it may be possible 
that even this is further decreased.

Now the two-$\Delta$ decay width into $NN\pi\pi$ is calculated 
as the double integral
\begin{equation}
\Gamma_{\rm 4,eff} = \frac{2}{\pi}\,  \frac{\int |\Psi_{\Delta\Delta}(p)|^2
[\Gamma(q_1) + \Gamma(q_2)]/2 \, p^2dp\, dq_1}
{q_{\rm max}\, \int_0^\infty |\Psi_{\Delta\Delta}(r)|^2\, r^2dr}
 \; .
 \label{double}
\end{equation}
Here the maximum limit of the free variable $p$ is obviously
from the kinematics of Fig. \ref{Deltas3}
$p_{\rm max} = \sqrt{s/4 - (M+\mu)^2}$ and the upper limit of
the pion momentum as a function of $p$ is obtained from the
maximum internal energy of particle one
\begin{equation}
s_{\rm 1max} = [\sqrt{s} - \sqrt{(M+\mu)^2+p^2}\;]^2 - p^2
\end{equation}
as
\begin{equation}
q_{\rm 1max}^2 =  \frac{(s_{\rm 1max} - M^2 - \mu^2)^2
- 4\mu^2M^2}{4\, s_{\rm 1max}} \; .
\end{equation}
In the pion integration the second dependent momentum $q_2$
in turn is obtained from
\begin{equation}
q_2^2 = \frac{(s_2 -M^2 -\mu^2)^2 - 4\mu^2M^2}{4\, s_2}
\end{equation}
with $s_2 = [\sqrt{s} - \sqrt{s_1+p^2}\;]^2 - p^2$ and
$s_1 = (M^2+q_1^2) + (\mu^2+q_1^2)$.

\begin{figure}[tb]
\centering
\sidecaption
\includegraphics[width=9cm]{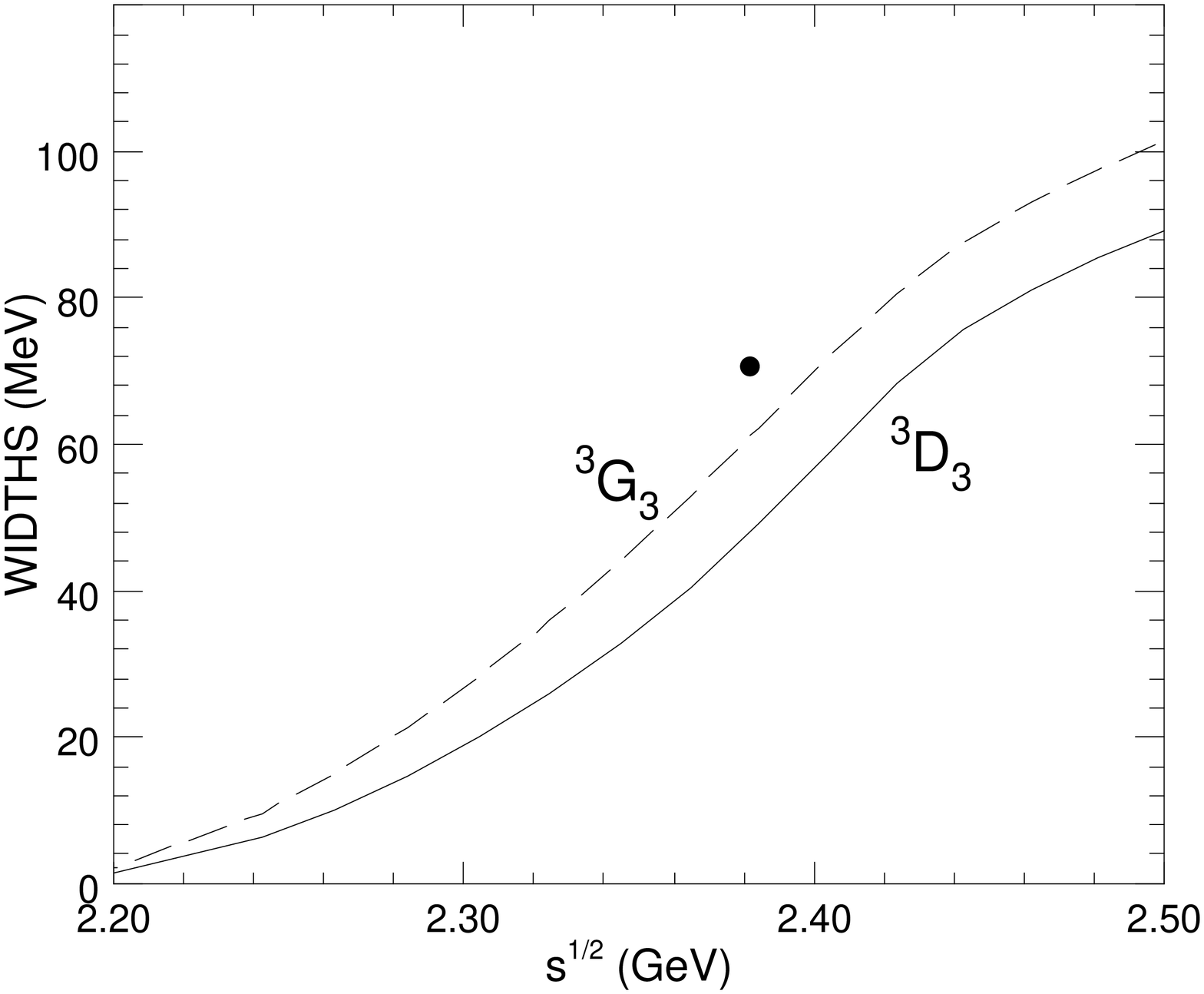}
\caption{The widths of the $^7S_3(\Delta\Delta)$ state in 
$I=0\;\; NN$ scattering: The solid curve arises from $^3D_3(NN)$
and the dashed one from $^3G_3(NN)$. The bullet shows the energy
and width of the resonance reported e.g. in Ref.~\cite{adlar90}.
\label{dgwid}
}
\end{figure}

Presently the $\Delta\Delta$ width in calculated for $NN$
scattering in isospin zero $3^+$ ({\it i.e.} $^3D_3$ and $^3G_3$)
state(s) reported as a resonance $d'(2380)$ discovered in
WASA@COSY experiments
\cite{adlar90} and speculated as a possible dibaryon.
The result is shown in Fig.~\ref{dgwid}. Both curves 
indicate a width less than a single free $\Delta$ and 
relatively well agreeing with the experimental value.
The narrowness may be considered surprising, though it is in
line with the results of Sec.~\ref{single} and also of
Gal and Garcilazo~\cite{galgar,gal}.

\section{Critique}
\label{critique}
The present calculation does not purport to be a genuine dynamic
{\it ab initio} field theory starting from interaction vertices.
Rather the ``vertices'' in Fig.~\ref{Deltas3}
illustrate the actual observable width 
for the decay $\Delta \rightarrow N\pi$. A deeper
work would involve also complex meson exchanges~\cite{arenbound}, 
which would 
probably increase inelasticity. On the other hand these may be 
strongly attractive and long-ranged bringing the mass from
the $\Delta\Delta$ threshold cusp at 2440 MeV down to the 
$d'(2380)$ region. Also inelasticities due to heavier
particles $\rho$ or $N'(1440)$, the Roper resonance, are absent.
Further, one might question the validity of the extension of the
parametrization of the experimental free width to the high
momenta needed in Eq.~\ref{double}.

In spite of these shortcomings and the lack of a quark calculation
here, the present results may throw some doubt on the inevitability
of $d'(2380)$ being the ``smoking gun'' manifestly demonstrating
quark exotics in intermediate energy $NN$ scattering based
mainly on its narrowness. 
A similar conclusion may be implied also from Ref.~\cite{galgar}.
An interesting extension of these calculations would be to isospin
2 or 3, which are not directly coupled with two-nucleon scattering 
states.

\end{document}